\documentstyle[aps,prb,twocolumn,epsf,floats]{revtex}
\begin{document}

\twocolumn[\hsize\textwidth\columnwidth\hsize\csname @twocolumnfalse\endcsname


\title{
Heat transport and spin-charge separation
in the normal state of high temperature superconductors}

\author{Qimiao Si}
\address{
Department of Physics and Astronomy, Rice University, Houston,
TX 77251-1892 }

\maketitle
\begin{abstract}

{
Hill et al. have recently measured both the thermal and charge
conductivities in the normal state of a high temperature superconductor.
Based on the vanishing of the Wiedemann-Franz ratio
in the extrapolated zero temperature limit, they conclude that
the charge carriers in this material are not fermionic. Here I make 
a simple observation that the prefactor in the
temperature dependence of the measured thermal conductivity
is unusually large, corresponding to an extremely small energy scale 
$T_0 \approx 0.15$ K. I argue that $T_0$ should be interpreted as 
a collective scale. Based on model-independent considerations,
I also argue that the experiment leads to two possibilities:
1) The charge-carrying excitations are non-fermionic. And much of the
heat current is in fact carried by distinctive charge-neutral
excitations; 2) The charge-carrying excitations are fermionic,
but a subtle ordering transition occurs at $T_0$.
}

\end{abstract}
\pacs{}]

Spin-charge separation has long been proposed to describe the normal state
of the high temperature superconductors\cite{Anderson1}, and 
continues to be the subject of extensive current work.
It has, however, received no unambiguous experimental support,
many indirect evidences notwithstanding\cite{Anderson2}.
One natural means to probe spin-charge separation is to compare
the temperature dependences of spin transport and charge
transport properties\cite{Si97,Si00,Fisher}. Spin injection experiments
have recently been carried out in the cuprates\cite{Goldman,Venkatesan,Yeh};
these experiments, however, have yet to yield quantitative information 
about spin transport. A less direct alternative to spin transport in this
context is heat transport, if the electronic contribution can be
unambiguously separated from the phononic one. While such a separation is
easy to achieve in highly conductive metals, it is in general very
difficult for strongly correlated metals\cite{Ong}.

Very recently, Hill et al.\cite{Taillefer} have extracted the electronic
contribution to the thermal conductivity ($\kappa$)
in the normal state
of an optimally electron-doped PCCO. This is achieved at very low
temperatures, 
where scattering is dominated by elastic processes. The normal state 
arises in a magnetic field applied along the $c-$axis
$H \approx 13~$T, which is above the bulk upper critical field $H_{c2}$.
(The thermal conductivity 
is field-independent at $H > H_{c2}$.) At roughly the same magnetic field,
the electrical conductivity ($\sigma$) is essentially temperature-independent,
up to small weak-localization-like corrections.
The authors take advantage of the fact that the
measured zero-field thermal conductivity $\kappa(H=0)$ goes to zero in 
approximately a $T^3$ fashion, in sharp contrast to what is generally
expected for a disordered d-wave superconductor\cite{Durst}.
They assume that $\kappa(H=0)$
is entirely due to phonons, and identify the electronic contribution 
to the normal-state thermal conductivity as 
$\kappa_e = \kappa(H \approx 13~{\rm T} ) - \kappa (H=0)$.
$\kappa_e$ is found to have an asymptotic low temperature form 
\begin{eqnarray}
	{\rm lim}_{T\rightarrow 0} ~\kappa_e \sim T^{\alpha+1}	
\label{L}
\end{eqnarray}
where $\alpha \approx 2.6$.
This is in strong violation of the Wiedemann-Franz (WF) law:
According to this law, 
$\alpha$ should be equal to zero reflecting the linear temperature
dependence of the specific heat of the fermionic quasiparticles 
in a Fermi liquid. From this perspective, the experiment implies 
a missing entropy. The authors conclude that the carriers of the charge
current are not fermionic.

Here I point out that the prefactor in the temperature dependence of the
thermal conductivity should be considered to be unusually large,
if the dominant contribution to $\kappa_e$ were due to non-fermionic
charge-carrying excitations.
To see this, 
we note that 
the measured $\kappa_e$ becomes larger than
\begin{eqnarray}
\kappa_{WF} \equiv L_0 T \sigma 
\label{kappa-WF}
\end{eqnarray}
at a temperature $T_0 \approx 0.15 K$. 
(Here $L_0 \equiv \pi^2 k_B^2 / 3 e ^2$ is the Lorenz number of a Fermi
liquid.) Namely, the experimental data can be cast in the form,
\begin{eqnarray}
  {\rm lim}_{T\rightarrow 0}~
{ \kappa_e \over \kappa_{WF}} \approx \left ( {T \over T_0} \right)^\alpha
\label{L2}
\end{eqnarray}
Using $\sigma = e^2 (d n / d \mu) D_{charge}$, 
where $d n / d \mu$ and $D_{charge}$ are the electronic
compressibility and charge diffusion constant, respectively,
we can rewrite 
\begin{eqnarray}
\kappa_{WF} = L_0 T (e^2 /m) n_{charge} \tau_0 
{{(d n / d \mu)} \over N_0} 
{{D_{charge} } \over D_0}
\label{kappa-WF-2}
\end{eqnarray}
where the subscript $0$ labels quantities in the
absence of interactions. 
It follows that $\kappa_{WF}$ does not explicitly depend on the
velocity and, equivalently, the bandwidth, of the charge-carrying
excitations. 

If the measured $\kappa_e $ is dominated by the contribution of the
charge-carrying excitations, we can express $\kappa_e / \kappa_{WF}$
in the following general form:
\begin{eqnarray}
{ \kappa_{e} \over 
\kappa_{WF} } = {C_v \over C_v^0 } 
{ N_0 \over {(d n / d \mu)}} {{D_{heat} } \over D_{charge}}
\label{WF-ratio}
\end{eqnarray}
where $D_{heat}$ is the entropy diffusion constant and $C_v$  the 
specific heat due to the charge-carrying excitations.
In this case, the scattering times for both the heat and charge currents
reflect the elastic scattering time of the same excitation.
As a result, the ratio $D_{heat} / D_{charge}$ is expected to depend
only on equilibrium interaction parameters\cite{note-inelastic}.
These interaction parameters should mostly cancel out in the product
${ {D_{heat} N_0} \over {D_{charge} (d n / d \mu) }}$.
(In the Fermi liquid theory with s-wave scattering,
this product 
is equal to $m/m^*$ making $\kappa_{e} \over \kappa_{WF}$
exactly equal to unity\cite{Castellani}.)
The temperature dependence of the ratio ${ \kappa_e \over \kappa_{WF}}$
then describes, in a dimensionless form, the temperature dependence of
the specific heat due to the charge-carrying excitations.
It is then clear that the prefactor of the asymptotic low-temperature
dependence of $\kappa_e$ should be 
considered to be anomalously large: The effective bandwidth of 
the charge-carrying excitations would be of order $0.15 K$.
Such a small bare scale is essentially impossible.

We are then forced to associate $T_0$ with a collective scale.
Unlike for a bare scale such as a bandwidth, it is entirely
reasonable for a collective scale to be this small.
On purely phenomenologically grounds,
there are two possible interpretations as illustrated in
Figs. 1a and 1b.

\begin{figure}[t!]
\centering
\vbox{\epsfysize=
8cm
\epsfbox{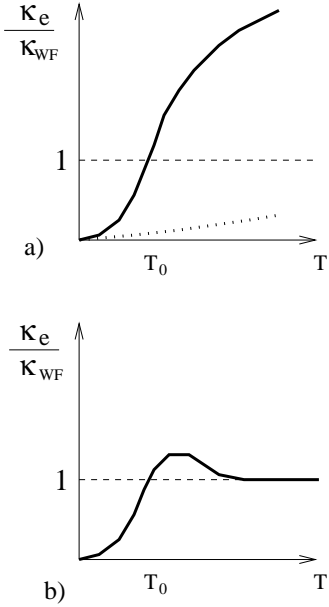}}
\vspace{2ex}
\caption{
Schematics of two possible pictures. The solid lines give the ratio of
the total electronic thermal conductivity, $\kappa_e$, to $\kappa_{WF}$,
defined in Eq. (\ref{kappa-WF}).
In a), the dotted line describes the contribution from
charge-carrying excitations; the remaining contribution comes from
charge-neutral excitations. In b), the excitations are fermionic and 
the low temperature behavior reflects an ordering transition around 
$T_0$.}
\label{fig1}
\end{figure}

{\em Distinctive charge-neutral excitations in the normal state:}
If the charge-carrying excitations indeed have non-fermionic statistics,
such that their contribution to the specific heat has a higher than linear 
temperature dependence, much of the heat current is necessarily carried by 
some distinctive charge-neutral excitations.
This conclusion is reached as follows. In this picture,
the contribution to the thermal conductivity due to
the charge-carrying excitations would have to be
\begin{eqnarray}
{ \kappa_{charge} \over \kappa_{WF}} \approx \left ( {T \over W} 
\right)^\beta
\label{L3}
\end{eqnarray}
where the exponent $\beta > 0$ reflects both the statistics and the
dispersion of the charge-carrying excitations, and $W$ is the effective
bandwidth. Since $W$ is expected to be much larger than $T_0$,
in the measured temperature range $\kappa_{charge}$ is necessarily 
much smaller than the measured $\kappa_e$ as illustrated in
Fig. 1a -- hence our conclusion. 
We note that a microscopic theory for such a picture needs to produce 
a temperature-independent carrier concentration $n_{charge}$ relevant
to transport and, at the same time, a specific heat that goes as
$T^{1+\beta}$.

Experimentally, the plot\cite{Taillefer} of $\kappa_e/T$ versus $T$
starts to deviate from the low temperature power-law form as 
the temperature is increased through $T_0$.
Within our picture, $T_0$ would be some collective temperature scale
associated with a transition that affects mostly the charge-neutral
excitations. It follows that the temperature dependence
of $\kappa_{neutral}$ is very different from that of $\kappa_{charge}$,
reflecting the separation of electrons into the charge-carrying and 
charge-neutral excitations.

Measurement of the heat transport cannot tell us about the quantum
number of the charge-neutral excitations. A natural candidate,
of course, would be that these are spin-carrying excitations.
Measurement of the spin transport would help clarify the nature of
these charge-neutral excitations.

{\em Ordering transition in the normal state:}
An alternative possibility is that, the charge-carrying excitations 
above $T_0$ are in fact fermions. A subtle ordering transition takes
place around $T_0$. Again, it is natural for $T_0$ to be so small since it is
a collective scale.

In this picture, the transition around $T_0$ leads to a sharp drop in the
specific heat but leaves the electrical conductivity largely unaffected.
In addition, the fermionic excitations are gapped out at $T << T_0$.
The reason that $\kappa_e/T$ goes to zero in the asymptotic low 
temperature limit is presumably related to what is responsible for 
the vanishing $\kappa(H=0)/T$.
While it is in principle possible, this picture, in its simple form,
requires some delicate balance between the quasiparticle contribution
and collective contribution to the charge current such that the 
electrical conductivity is essentially unchanged as the temperature is
lowered through $T_0$. The collective contribution is of relevance here since
disorder is not very strong and pinning is expected to be weak.
Whether such a balance can be achieved in specific microscopic models
remains to be seen.

Experimentally, $\kappa_e/T\sigma$ at temperatures above $T_0$ 
is indeed close to\cite{Lee}, though somewhat larger than, $L_0$.

One way to differentiate these two pictures is to study the behavior
of $\kappa_e$ at temperatures high compared to $T_0$
(but still low enough so that elastic scatterings dominate).
In the first picture, the mean free path of the charge-neutral excitations
is in general very different from that of the charge-carrying
excitations. As illustrated in Fig. 1a,
we expect $\kappa_e / \kappa_{WF}$ to be very different
from unity for $T >> T_0$.
In the second picture, on the other hand, we expect
$\kappa_e / \kappa_{WF} \approx 1 $ for $T >> T_0$; see Fig. 1b.
Unfortunately, the experiment in PCCO is limited to temperatures
not too high compared to $T_0$.
Analogous experiments in other cuprates may help clarify the
situation.

I would like to thank L. Taillefer for sending me a copy of
Ref. \onlinecite{Taillefer} prior to publication,
and E. Abrahams, A. J. Leggett, L. Taillefer, and C. M. Varma 
for useful discussions. 
This work was carried out during my participation in the 
ITP-UCSB program on high temperature superconductivity;
I would like to thank the organizers of this program and ITP
for hospitality. The work has been supported in part by NSF Grant
No. PHY99-07949 (ITP), NSF Grant No. DMR-0090071, and TCSUH.


\begin{thebibliography}{9}
\bibitem{note-inelastic} Here we consider elastic scatterings only.
For inelastic scatterings, there are exceptions to this statement.
\end{thebibliography}
\end{document}